\documentclass{aa} 
\usepackage{times,graphicx,epsfig,color}
\usepackage[colorlinks=true,allcolors=blue]{hyperref}
\newcommand{\bq}{\begin{equation}}
\newcommand{\eq}{\end{equation}}
\newcommand{\bqn}{\begin{eqnarray}}
\newcommand{\eqn}{\end{eqnarray}}
\usepackage[flushleft]{threeparttable}
\usepackage{natbib}
\bibpunct{(}{)}{;}{a}{}{,} 
\begin{document}
\title{The local rotation curve of the Milky Way based on SEGUE and RAVE data}

\author{
K. Sysoliatina$^{1}$\thanks{Fellow of the International Max Planck Research School for Astronomy
and Cosmic Physics at the University of Heidelberg (IMPRS-HD).}, 
A. Just$^{1}$,
O. Golubov$^{2,3,4}$, 
Q.A. Parker$^{5,6}$,
E.K. Grebel$^{1}$,
G. Kordopatis$^{17}$,
T. Zwitter$^{15}$,
\mbox{J. Bland-Hawthorn$^{14}$},
B.K. Gibson$^{8}$,
A. Kunder$^{16}$,
U. Munari$^{7}$,
J. Navarro$^{12}$,
W. Reid$^{10,11}$,
G. Seabroke$^{9}$,
M.Steinmetz$^{18}$, and
F. Watson $^{13}$
}

\institute{
$^{1}$Astronomisches Rechen-Institut, Zentrum f\"{u}r Astronomie der Universit\"{a}t Heidelberg,
M\"{o}nchhofstr. 12--14, 69120 Heidelberg, Germany\\
$^{2}$Schools of Physics and Technology, Karazin Kharkiv National University, 4 Svobody Sq., Kharkiv, 61022, Ukraine\\
$^{3}$Institute of Astronomy, Karazin Kharkiv National University, 35 Sumska Str., Kharkiv, 61022, Ukraine\\
$^{4}$Department of Aerospace Engineering Sciences, University of Colorado at Boulder, 429 UCB, Boulder, CO, 80309, USA\\
$^{5}$Department of Physics, the University of Hong Kong, Hong Kong SAR, China \\
$^{6}$The Laboratory for Space Research, the University of Hong Kong, Hong Kong SAR, China\\
$^{7}$INAF Astronomical Observatory of Padova, 36012 Asiago (VI), Italy\\
$^{8}$E.A. Milne Centre for Astrophysics, University of Hull, Hull, HU6 7RX, UK\\
$^{9}$Mullard Space Science Laboratory, University College London, Holmbury St Mary, Dorking, RH5 6NT, UK\\
$^{10}$Department of Physics and Astronomy, Macquarie University, Sydney, NSW 2109, Australia\\
$^{11}$Western Sydney University, Locked bag 1797, Penrith South DC, NSW 2751, Australia\\
$^{12}$Senior CIfAR Fellow, University of Victoria, Victoria BC, Canada V8P 5C2\\
$^{13}$Australian Astronomical Observatory, PO Box 915, North Ryde, NSW 1670, Australia\\
$^{14}$Sydney Institute for Astronomy, School of Physics, University of Sydney, NSW 2006, Australia\\
$^{15}$Faculty of Mathematics and Physics, University of Ljubljana, 1000 Ljubljana, Slovenia\\
$^{16}$Saint Martin's University, 5000 Abbey Way, Lacey, WA 98503, USA\\
$^{17}$Universit\'{e} C\^{o}te d'Azur, Observatoire de la C\^{o}te d'Azur, CNRS, Laboratoire Lagrange, France\\
$^{18}$Leibniz Institut f\"{u}r Astrophysik Potsdam, An der Sternwarte 16, D-14482, Potsdam, Germany 
}

\date{Printed: \today}

\abstract
{} 
{We construct the rotation curve of the Milky Way in the extended solar neighbourhood using a sample of SEGUE (Sloan Extension for Galactic Understanding and Exploration) G-dwarfs. 
We investigate the rotation curve shape for the presence of any peculiarities just outside the solar radius as has been reported by some authors.} 
{Using the modified Str\"{o}mberg relation and the most recent data from RAVE (RAdial Velocity Experiment), 
we determine the solar peculiar velocity and the radial scalelengths for the three populations of different metallicities representing the Galactic thin disc. 
Then with the same binning in metallicity for the SEGUE G-dwarfs, we construct the rotation curve in the range of Galactocentric distances \mbox{$7-10$ kpc}. 
We approach this problem in a framework of classical Jeans analysis and derive the circular velocity 
by correcting the mean tangential velocity for the asymmetric drift in each distance bin. 
With SEGUE data we also calculate the radial scalelength of the thick disc taking as known the derived peculiar motion of the Sun and the slope of the rotation curve.} 
{The tangential component of the solar peculiar velocity is found to be $V_{\odot}=4.47\pm 0.8$ km s$^{-1}$ and the corresponding scalelengths 
from the RAVE data are \mbox{$R_d$(0$<$[Fe/H]$<$0.2$)=2.07\pm0.2$} kpc, \mbox{$R_d$(-0.2$<$[Fe/H]$<$0$)=2.28\pm0.26$} kpc 
and \mbox{$R_d$(-0.5$<$[Fe/H]$<$-0.2$)=3.05\pm0.43$ kpc}. 
In terms of the asymmetric drift, the thin disc SEGUE stars are demonstrated to have dynamics similar to the thin disc RAVE stars,
therefore the scalelengths calculated from the SEGUE sample have close values: 
\mbox{$R_d$(0$<$[Fe/H]$<$0.2$)=1.91\pm0.23$} kpc, \mbox{$R_d$(-0.2$<$[Fe/H]$<$0$)=2.51\pm0.25$} kpc and \mbox{$R_d$(-0.5$<$[Fe/H]$<$-0.2$)=3.55\pm0.42$ kpc}.
The rotation curve constructed through SEGUE G-dwarfs appears to be smooth in the selected radial range \mbox{7\,kpc $< R <$ 10\,kpc}.
The inferred power law index of the rotation curve is $0.033\pm 0.034$, which corresponds to a local slope of $dV_c/dR=0.98\pm 1$ km s$^{-1}$ kpc$^{-1}$.
The radial scalelength of the thick disc is 2.05 kpc with no essential dependence on metallicity. 
}
{The local kinematics of the thin disc rotation as determined in the framework of our new careful analysis 
does not favour the presence of a massive overdensity ring just outside the solar radius. 
We also find values for solar peculiar motion, radial scalelengths of thick disc and three thin disc populations of different metallicities as a side result of this work.
}


\titlerunning{The local rotation curve of the Milky Way based on SEGUE and RAVE data}
\authorrunning{K. Sysoliatina et al.}

\offprints{K. Sysoliatina}
\mail{Sysoliatina@uni-heidelberg.de}


\maketitle


\section{Introduction\label{intro}}

The measurement of the Galactic rotation curve provides a powerful tool for constraining the mass distribution 
in the Milky Way and enters various branches of Galactic kinematics as an essential ingredient.

The measurement of the rotation curve inside the solar orbit at $R_0$ can be done without even knowing the distances to the tracers. 
Spectroscopic observations of HI regions and molecular clouds emitting in the radio range yield their line-of-sight velocities, 
which under the assumption of the circularity of orbits can be converted to the circular velocities purely geometrically. 
Though this technique known as the tangent point method (TPM, \mbox{\citealp{binney}}) provides a quite accurate measurement of the inner rotation curve,
it loses its applicability (1) at small Galactocentric distances $R \, < \, 5$ kpc, where the bar starts to dominate 
(however, see \mbox{\citealp{wegg15}} for a recent long bar model)
and therefore the orbits can significantly deviate from 
circular ones (\citealp{sofue}) and (2) in the outer disc for $R \, > \, R_0$, where the distances to the tracers cannot be derived from a simple geometry. 
And even in the 'good' range of Galactocentric distances $5\,\mathrm{kpc}<R<R_0$, the reliability of the TPM may 
be questionable as was shown by recent studies, which consider the spiral structure in galactic discs \mbox{\citep{chemin15,chemin16}}. 
In order to probe the outer rotation curve one is obligated to determine distances to the tracers. The distance uncertainties together with the decrease of 
tracers' density with increase of R explain why the outer rotation curve is known less confidently than its inner part. 
However, very long baseline interferometry (VLBI) provides high-accuracy measurements of parallaxes and proper motions of young star-forming regions and  
covers a broad range of Galactocentric distances giving a strong constraint on the shape of the rotation curve \citep{reid14}. 
Still, due to the variety of techniques, difficulties in obtaining the accurate 6D dynamical information on coordinates and velocities, change in methodology 
at $R=R_0$ if one relies on the TPM inside the solar radius, the rotation curves determined by different authors are not in a perfect agreement with each other \citep{bhawthorn}.

\mbox{\citet{sofue}} presented a comprehensive analysis of previous measurements of the outer rotation curve, 
including data for HII regions and C stars, as well as points obtained by the HI-disc thickness method and VLBI observations.
The authors claimed a dip in the rotation curve between 7 and 11 kpc from the Galactic centre, which they attribute 
to the presence of a ring of stellar overdensity in the Galactic disc influencing the gravitational potential.
The dip is centred at 9 kpc, where the circular velocity drops by \mbox{$\sim$15 km s$^{-1}$}. 
\mbox{\citet{huang16}} also found a similar depression in the rotation curve obtained from stars of the Sloan Digital Sky Survey III's 
Apache Point Observatory Galactic Evolution Experiment (SDSS/APOGEE, \mbox{\citealp{eisenstein11}}) and the Large Sky Area Multi-Object Fiber Spectroscopic Telescope (LAMOST)
Spectroscopic Survey of the Galactic Anti-centre (LSS-GAC, \mbox{\citealp{liu14}}).
\mbox{\citet{kafle12}} used blue horizontal branch stars to construct a rotation curve, which also has a dip, although at bigger radii, about 11 kpc.
In contrast to these results, \mbox{\citet{lopez14}}, who studied proper motions of disc red clump giants, obtained a flat rotation curve without any dip, 
although the errors still remain substantial.
Having analysed a sample of APOGEE stars covering the range of Galactocentric distances 4-14 kpc, \mbox{\citet{bovy12a}}
arrived at the same conclusion - the authors found that the rotation curve is approximately flat. 
Another flat rotation curve was obtained by \mbox{\citet{reid14}} by studying high-mass star-forming regions.

These apparently contradictory results show that a more detailed study of the local Galactic rotation curve is strongly warranted. Now, with the abundant data for distances and 
velocities of millions of stars from photometric and spectroscopic surveys of the last decade, the situation is more encouraging. 
However, as we show in this paper, the solution of this task is not straightforward: while deriving the mean rotation velocity from the kinematic data,
one has carefully account for the asymmetric drift correction, which itself requires some knowledge or plausible assumptions about the Galactic potential. 
Because of this inter-dependence between the input and output we have to approach the problem of the rotation curve reconstruction in an iterative and consistent way.

This paper is organized as follows. Section \ref{samples} describes the data samples and the selection criteria. Section \ref{Jeans} contains the basics 
of our analytic approach in the most general form. Then we apply it in two consequent steps. First, we investigate the peculiar motion of the Sun 
with the local data from the RAdial Velocity Experiment (RAVE, \mbox{\citealp{steinmetz06}}). This analysis is presented in \mbox{Section \ref{RE}}. 
In the second step in \mbox{Section \ref{RC}} we construct the rotation curve of the extended solar neighbourhood using the 
sample of G-dwarfs from the Sloan Extension for Galactic Understanding and Exploration (SEGUE, \citealp{yanny09}).
In Section \ref{conclusions} we summarize our findings, discuss dependencies on the assumed constants and parameters and conclude.
Our treatment of the tilt term of the velocity ellipsoid is discussed in Appendix \ref{tracer}.

\section{Data samples\label{samples}}

\begin{figure*}
\centerline{\resizebox{\hsize}{!}{\includegraphics{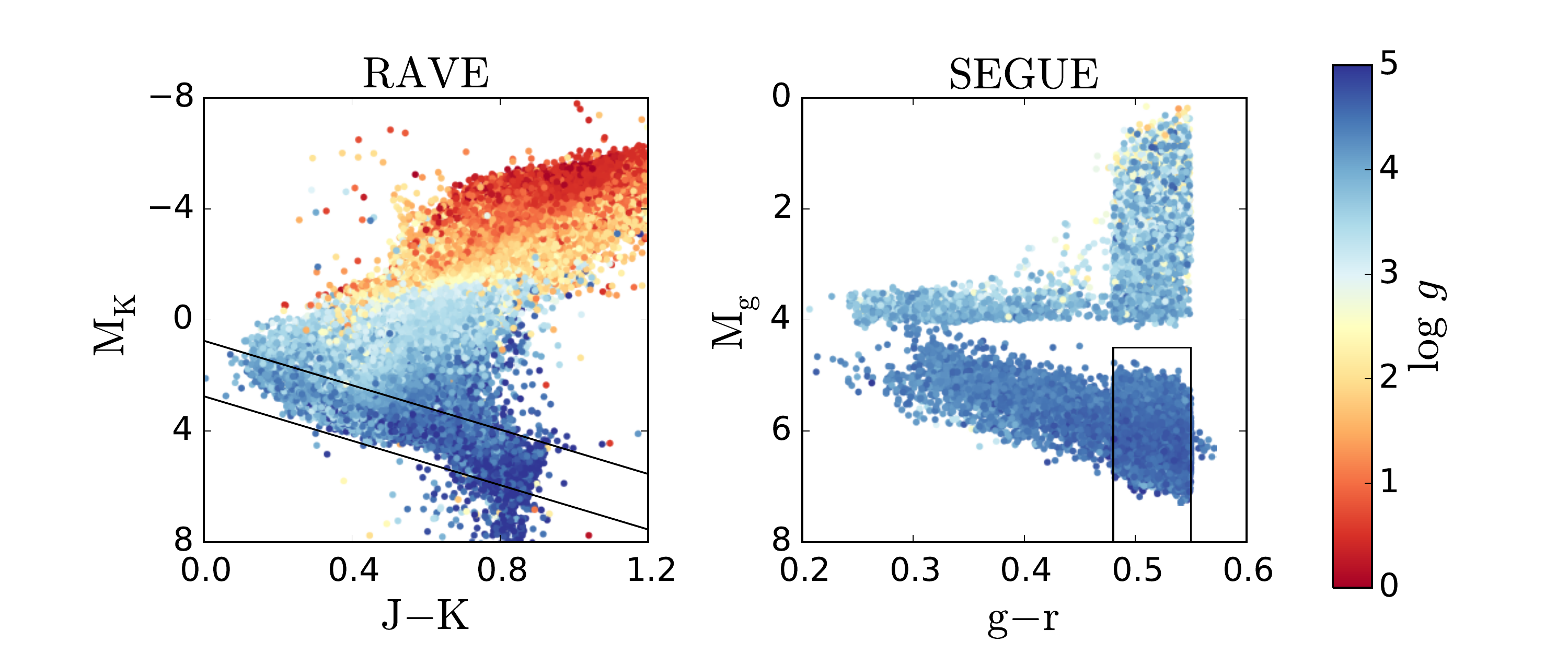}}}
\caption{The colour-magnitude diagrams of the whole RAVE and SEGUE G-dwarf data 
from 2MASS and SDSS photometry, respectively. The applied cuts are shown in black lines.}
\label{cmds}
\end{figure*}

\begin{figure}
\centerline{\resizebox{\hsize}{!}{\includegraphics{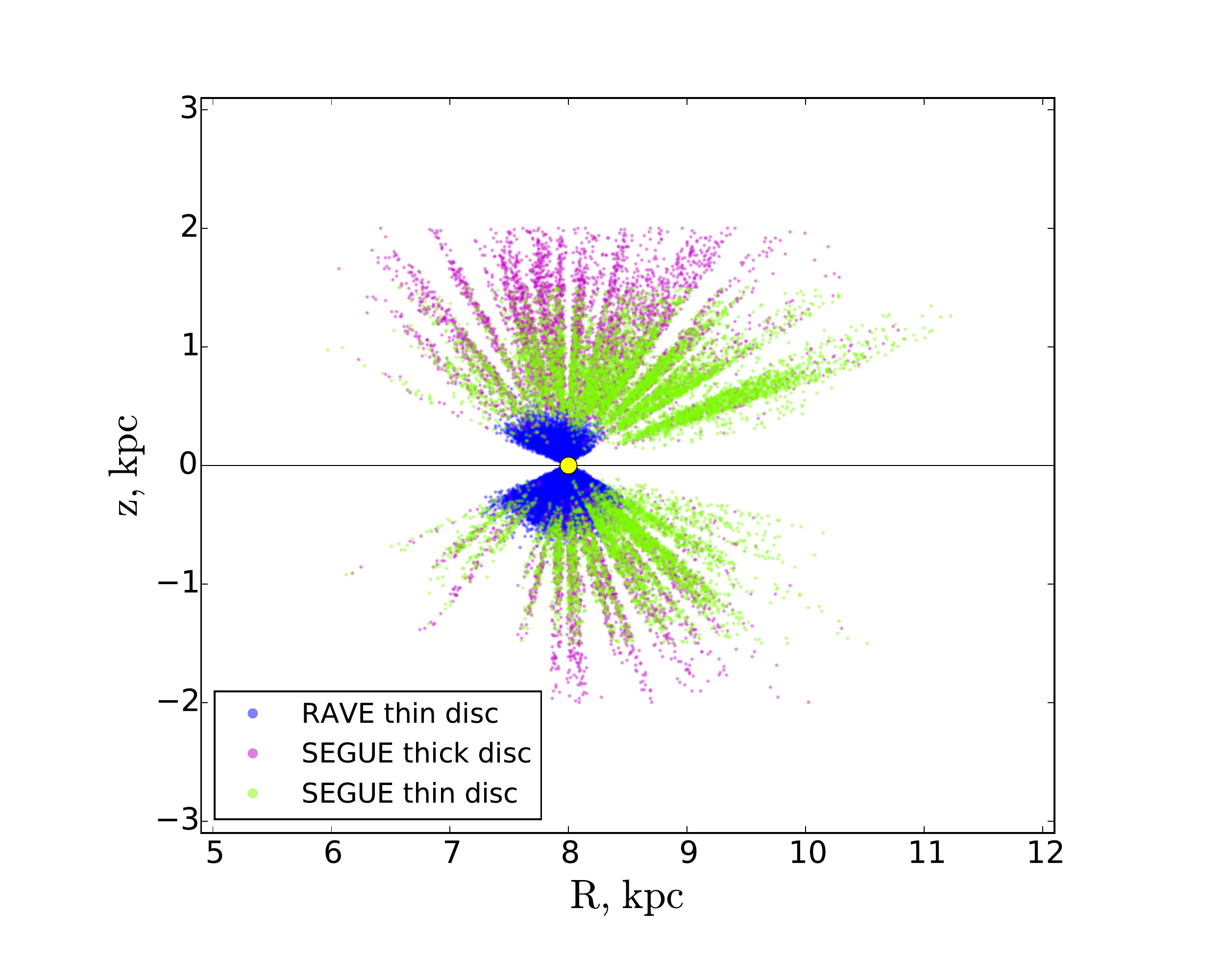}}}
\centerline{\resizebox{\hsize}{!}{\includegraphics{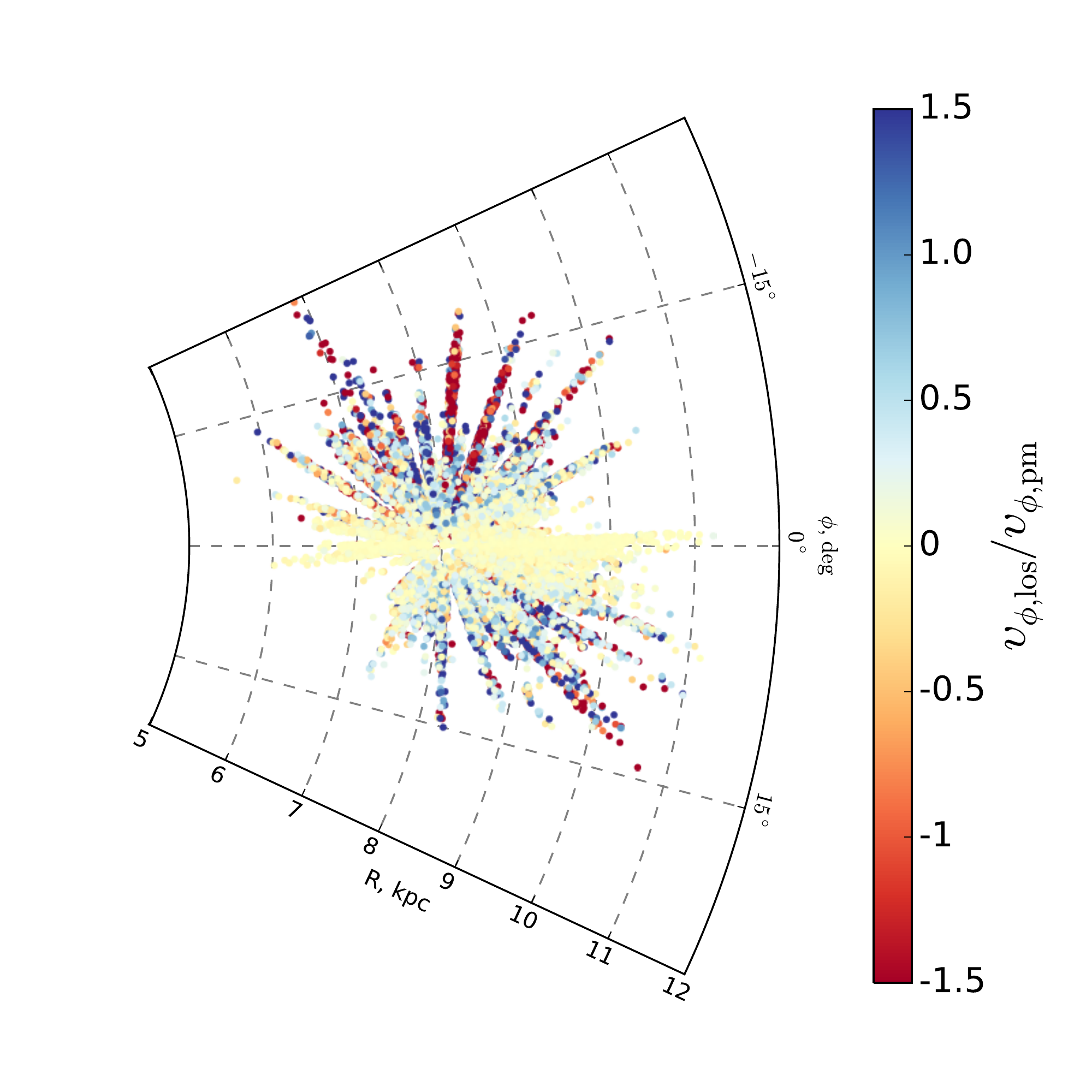}}}
\caption{The spatial distribution of the RAVE and SEGUE samples.  
The top panel shows Galactic cylindrical coordinates $R$ and $z$ of the final SEGUE (both thin and thick disc stars) and RAVE data samples (only thin disc stars).
The position of the Sun at Galactocentric distance \mbox{$R_0=8$ kpc} and height \mbox{$z=0$ kpc} is marked with a yellow circle.
The bottom panel shows the whole SEGUE sample projected on the Galactic plane. The value of $\phi=0^\circ$ corresponds to the Sun-Galactic centre axis. 
}
\label{samples-plot}
\end{figure}

Our analysis is performed in two steps. Firstly, we re-analyse the determination of the peculiar motion of the Sun and the radial scalelengths 
of the three thin disc populations in the framework of the approach of \mbox{\citet{golubov13}}, but based on the most recent data release of RAVE 
(DR5, \mbox{\citealp{kunder17}}) and a more careful treatment of the asymmetric drift correction (see \mbox{Section \ref{RE}}). 
RAVE is a kinematically unbiased spectroscopic survey with medium resolution (R$\sim$7500), 
which provides line-of-sight velocities, stellar parameters, and element abundances for more than 520~000 stars.
In RAVE DR5 improved stellar parameters and abundances were published and \mbox{\citet{McMillan17}} derived new distances taking into account 
the parallaxes of the Tycho-Gaia Astrometric Solution (TGAS) from the first Gaia data release (Gaia DR1, \mbox{\citealp{gaia16}}). 
Together with the proper motions from UCAC5 (The fifth US Naval Observatory CCD Astrograph Catalog, \mbox{\citealp{zacharias17}}) provided for most of the RAVE DR5 stars, 
this sample constitutes the basis of the most reliable local kinematic dataset to date. 
From this sample we select stars at Galactic latitudes $|b|>20^{\circ}$ (to avoid the necessity to consider extinction), 
belonging to the stripe with $0.75<K-4(J-K)<2.75$ on the colour-magnitude diagram (CMD) constructed with 2 Micron All Sky Survey photometry (2MASS, \citealp{skrutskie06}) 
(to select the main sequence and have a more uniform population of stars; \mbox{Figure \ref{cmds}}, left panel), 
with a signal-to-noise ratio S/N$\geq$30, relative distance errors $\delta d/d<0.5$, errors in proper motions $\delta\mu<10$ mas yr$^{-1}$  
and line-of-sight velocities $\delta V_{los}<3$ km s$^{-1}$. To select a cleaner thin disc sample we also use RAVE abundances and take only stars with [Mg/Fe]$<$0.2 \mbox{\citep{wojno16}}. 
Our final RAVE sample contains 23~478 stars. 
Being very local (Figure~\ref{samples-plot}, top panel), RAVE data cannot be directly employed for the reconstruction of the rotation curve, 
but rather are used for the purpose of investigating the solar peculiar motion. 

In the second step we use a sample of G-dwarfs from SEGUE, a low-resolution (R$\sim$2000) spectroscopic sub-survey of SDSS.  
The sample contains 40~496 G-dwarfs with photometric parallaxes measured by \mbox{\citet{lee11}} with a better than 10-\% accuracy for individual stars.
For our analysis we select stars with signal-to-noise ratio S/N$\geq$30 (to ensure good accuracy of the spectroscopic data), 
colour index $0.48\leq g-r\leq 0.55$ (to have a more uniform sample; most stars from the initial sample belong to this colour range anyway), 
and absolute magnitude $M_g>4.5$\,mag (which when combined with the colour cut, neatly selects the main sequence; Figure \ref{cmds}, right panel).
We separate the sample into the thin disc with [$\alpha$/Fe]$<$0.25, $|z|<1.5$ kpc and the thick disc with [$\alpha$/Fe]$>$0.25, $|z|<2$ kpc and [Fe/H]$>$-1.2 
(to decrease contamination by halo stars).
After applying all the selection criteria, 10~700 stars remain in the thin disc, and 7~040 stars in the thick disc samples.
SEGUE data extend over the range of Galactocentric distances of 7-10 kpc (Figure~\ref{samples-plot}), which makes them suitable for 
the local rotation curve analysis. 

The tangential velocities $\upsilon_{\phi}$ of the SEGUE stars rely on measurements of line-of-sight velocities and proper motions with distances to individual stars. 
Radial velocities and proper motions are obtained by different observational techniques, so they have different 
accuracy (about 3 km s$^{-1}$ for line-of-sight velocities and 3-4 mas yr$^{-1}$ for proper motions, which at a distance of 2 kpc from the Sun 
converts to a $\sim$ \mbox{30-40 km s $^{-1}$} error)
and might have essentially different systematics. The relative contributions of line-of-sight velocity $\upsilon_{\phi,los}$ and proper motion $\upsilon_{\phi,pm}$ terms to 
the resulting tangential velocity $\upsilon_{\phi}$ change with Galactocentric longitude, 
so it is important to check that our data when binned in Galactocentric distance are not entirely dominated by one of the terms. The bottom panel of Figure \ref{samples-plot}
shows the SEGUE sample (both thin and thick disc stars selected with our criteria) projected on the Galactic plane.
We calculate and compare the contributions $\upsilon_{\phi,los}$ and $\upsilon_{\phi,pm}$ measured relative to the solar Galactocentric velocity $v_\odot$. 
The contribution of the $\upsilon_{\phi,los}$ term is negligible on the axis Sun-Galactic centre, and in other regions values of the ratio reflect an interplay between the direction
and the speed of stellar motions. Importantly, all longitudes are represented in our sample equally well, so in the useful range of Galactocentric distances of 7-10 kpc we expect no bias in $\upsilon_{\phi}$ 
with respect to the observational techniques. 

It's also important to note here that the RAVE data are expected to be kinematically unbiased by construction (see \mbox{\citealp{wojno17}}), 
so we do not need to worry about selection effects while working with them. As to the SEGUE \mbox{G-dwarfs}, simple selection criteria were used 
to construct this survey \mbox{\citep{yanny09}}, so we expect these data to be relatively free of kinematic bias and therefore we do not include a correction for the selection effects for SEGUE as well. 
But as a measure of precaution, before using SEGUE data for the rotation curve reconstruction, we check our RAVE and SEGUE samples for consistency in terms of kinematics 
in order to justify our approach (see Section \ref{RE}).

\section{Jeans analysis\label{Jeans}}

In this section we discuss the properties of the asymmetric drift, when applied to a large volume in Galactocentric distance $R$ and 
height above the Galactic midplane $z$. 
The equation that we are going to formulate here is the basis of our work. 

The asymmetric drift is defined as the lag in tangential speed of tracer populations with respect to the rotation curve, $V_a = v_c - \overline{v_{\phi}}$. 
The exact value of this quantity depends on stellar population properties and varies with the position in the Galaxy. This implies 
that in order to convert mean tangential velocities $\overline{v_{\phi}}$ to the rotation curve, we need to correct them for 
$V_a$.
To quantify the asymmetric drift we use the same notation as in \mbox{\cite{golubov13}} and start with the radial Jeans equation for a stationary and axisymmetric system 
(in cylindrical coordinates and with negligible mean radial and vertical motion):
\begin{eqnarray}
v_c^2 - \overline{v_{\phi}}^2 &=& -R 
\left ( \sigma_{R}^2 \frac{\partial \ln(\nu \sigma_R^2)}{\partial R}
+ \frac{\sigma_R^2 -\sigma_{\phi}^2}{R} 
\right.\nonumber\\ &&\qquad\left.
+ \sigma_{Rz}^2 \frac{\partial \ln(\nu \sigma_{Rz}^2)}{\partial z} + F(R,z)
\right ).
\label{JE1}
\end{eqnarray}
Here $\nu$ and $\sigma^2_{R,z,\phi,Rz}$ are the tracer density and velocity ellipsoid, 
$\overline{v_{\phi}}$ is the mean tangential speed of the population, 
$v_c$ is its circular speed defined in the midplane and $F(R,z)$ measures the vertical variation of the radial force, i.e.,
\begin{equation}
F(R,z) \equiv \left.\frac{\partial \Phi}{\partial R} \right|_z - \left.\frac{\partial \Phi}{\partial R} \right|_0 
\quad;\quad v_c^2 = R \left.\frac{\partial \Phi}{\partial R} \right|_{0},
\label{ver_def}
\end{equation}
where $\Phi$ is the total gravitational potential. The right-hand side of Eq. \ref{JE1} is a measure of the asymmetric drift we are interested in.
The last term vanishes at $z=0$, but it should not be neglected at $z\neq0$. 
Indeed, as with the increase of height above the midplane the radial gradient of the Galactic gravitational potential decreases, 
so does the measured tangential speed.  
This effect, if not taken into account, can result in two biases. 
Firstly, the derived circular velocity could be underestimated by a few kilometres per second, 
causing a shift of the rotation curve as a whole. Secondly, the typical distance of stars from the midplane varies 
at different Galactocentric radii due to the sample geometry (see SEGUE sample in Figure \ref{samples-plot}).
This can cause a distortion of the rotation curve, which is unacceptable as the robust reconstruction of the local rotation curve's shape is the very aim of this work. 
The second last term is the well-known tilt term, which does not vanish in general even in the midplane (see Eq. \ref{tilt-term} below).
Since these vertical correction terms lead to 
a systematic variation of the asymmetric drift increasing with $|z|$, we will take them into account everywhere throughout this work, 
even for the relatively local RAVE sample with the range of useful distances limited to $\pm$ 0.5 kpc from the Sun (\mbox{Figure \ref{samples-plot}}, top panel).
The details of the derivation and the assumptions made are provided in \mbox{Appendix \ref{tracer}}.
Here we explain only the essential part of our treatment of \mbox{Eq. \ref{JE1}}.

\begin{table}[b]
\centering

\begin{threeparttable}
\caption{The properties of the Galaxy used for calculations.} 
\begin{tabular}{l|l|l} \hline
Variable & Value & Source \\ \hline
$\rho_{h\odot}$ ($M_\odot$ pc$^{-3}$) & 0.014 & (1) \\
$a_{h}$ (kpc) & 25 & \ -- \\ \hline
$M_b$ ($M_\odot$) & 1.1 $10^{10}$  & \ -- \\ \hline
$\Sigma_d$ ($M_\odot$ pc$^{-2}$) & 30 & (1) \\
$h_d$ (pc) & 300 & (1) \\
$R_d$ (kpc)& 2.5 & (2) \\ \hline
$\Sigma_t$ ($M_\odot$ pc$^{-2}$) & 6  & (3) \\
$h_t$ (pc) & 800 & (3) \\
$R_t$ (kpc) & 1.8 &  (4) \\ \hline
$\Sigma_g$ ($M_\odot$ pc$^{-2}$) & 10 & (1) \\
$h_g$ (pc) & 100 & (1) \\
$R_g$ (kpc) & 4.5 & (5) \\
$R_{cut,g}$ (kpc) & 4.0 &  \ -- \\
\hline
\end{tabular}

\begin{tablenotes}
 \item References. (1) \citealp{just10}; (2) \citealp{golubov13}; (3) \citealp{just11}; (4) \citealp{cheng12}; (5) \citealp{robin3}. 
\end{tablenotes}

\end{threeparttable}
\label{tab-var}
\end{table}

The rotation curve, which we wish to determine, depends on the same Galactic potential. Therefore, we need to check carefully that we are not 
biasing the result implicitly by adopting a special model for the potential, entering \mbox{Eq. \ref{JE1}} through the vertical gradients of tracer density and the radial force.
We use a five-component model of the Galaxy with a spherical Navarro-Frenk-White (NFW) Dark Matter (DM) halo \mbox{\citep{nfw}}
with a local density $\rho_{h\odot}$ and power law slope $\gamma_h=-1$, 
a bulge component with mass $M_b$ and three exponential discs for the gas, and the thin and thick disc with local surface densities $\Sigma_i$ and 
radial and vertical scalelengths $R_i$ and $h_i$. The gaseous disc has an inner hole with the radius of $R_{cut,g}=4$ kpc. 
The values of all used parameters are given in \mbox{Table \ref{tab-var}}. 
With this Galactic model we calculate the exact value of 
the vertical variation of the radial force using the $GalPot$ code\footnote{Developed by P.McMillan and available at \newline \url{https://github.com/PaulMcMillan-Astro/GalPot}} 
(a stand-alone version of Walter Dehnen's GalaxyPotential C++ code, \mbox{\citealp{dehnen98a}}). 

In the tilt term in Eq. \ref{JE1} we parametrize $\sigma_{Rz}^2$ as 

\begin{equation}
\sigma_{Rz}^2 = \eta (\sigma_R^2 - \sigma_z^2) z/R,
\label{sig_rz}
\end{equation}
where the parameter $\eta$ describes the orientation of the velocity ellipsoid relative to the Galactic centre direction.
The logarithmic derivative of $\eta\nu (\sigma_R^2-\sigma_z^2)$ can be further parametrized with some characteristic scaleheight $h_{\nu \sigma}$ 
(which describes the vertical variation of the tracer density and velocity ellipsoid orientation and is in general a function of $R$ and $z$), leading to
\begin{equation}
\sigma_{Rz}^2 \frac{\partial \ln (\nu \sigma_{Rz}^2)}{\partial z} =  
\eta \frac{\sigma_R^2 - \sigma_z^2}{R}\left[ 1 - \frac{z}{h_{\nu \sigma}}\right].
\label{tilt-term}
\end{equation}

The first term in Eq. \ref{JE1} can be characterized by the radial scalelength $R_E$ (which again may depend on $R$ and $z$), so it reads
\begin{equation}
\sigma_R^2 \frac{\partial \ln(\nu \sigma_R^2)}{\partial R} = - \frac{\sigma_R^2}{R_E}. 
\label{term5}
\end{equation}
At this point we remark that under the assumption of a constant disc thickness and a constant shape of the velocity ellipsoid 
$\sigma_z^2/\sigma_R^2 $, which implies $\nu \propto \sigma_z^2 \propto \sigma_R^2$, we find $R_E$ related to the radial scalelength of the tracer density 
$\nu$ through \mbox{$R_\nu = 2\,R_E = const.$} We will use this assumption later in order to convert the measured $R_E$ into the radial scalelengths 
of the subpopulations (see Section \ref{RE}). 

Finally, we can write the Jeans equation (Eq. \ref{JE1}) as
\bqn
v_c^2 - \overline{v_{\phi}}^2 &=& \sigma_R^2\left( \frac{R}{R_E}-1\right)  +\sigma_{\phi}^2 \nonumber\\
&& -R\, F(R,z) - \eta \left(\sigma_R^2 - \sigma_z^2\right)\left( 1 - \frac{z}{h_{\nu \sigma}}\right).
\label{JE2}
\eqn
This equation is still equivalent to Eq. \ref{JE1}. With a specification of the parameters/functions $F(R,z)$,
$\eta$, $h_{\nu \sigma}$, $R_E$ (\mbox{e.g. $R_E$} independent of $R,z$) we lose generality.

\section{Solar motion and radial scalelengths\label{RE}}

Before using the mean measured relative velocities of the tracer stars to find the mean tangential velocities $\overline{v_{\phi}}$ and 
correct them for the asymmetric drift, 
we need to correct for the peculiar motion of the Sun $(U,V,W)_\odot$, i.e., to convert the velocities to the local standard of rest.
The determination of $U_{\odot}$ and $W_{\odot}$ from the stellar kinematics in the solar neighbourhood is not a matter of difficulty. 
For the radial and vertical components of the solar motion we adopt the values from \mbox{\citet{schoenrich10}} 
\mbox{$U_\odot=11.1$ km s$^{-1}$}, \mbox{$W_\odot=7.25$ km s$^{-1}$}, which are also consistent with our data sets. 
In contrast, the determination of the tangential component $V_{\odot}$ is challenging. 
Due to the fact that the observed tangential motion of stellar populations is affected both
by the solar peculiar velocity and the asymmetric drift, the task of disentangling them poses a problem. 
The classical value based on \textit{Hipparcos} data is $V_\odot=5.2$ km s$^{-1}$ \mbox{\citep{dehnen98}}. The more recent and widely used ones 
lie approximately between 10 and 12 km s$^{-1}$ (e.g., $12.24 \pm 0.47$ km s$^{-1}$ in \mbox{\citealp{schoenrich10}}). However, among recent estimates there are also 
values as high as $\sim$24 km s$^{-1}$ \mbox{\citep{bovy12}} indicating that the local stellar motions could be also influenced by non-axisymmetric Galactic features 
like spiral arms \citep{siebert12,monari16} and a bar \citep{dehnen00,monari17,perez-villegas17}.
At the lower limit there is $V_{\odot}=3.06 \pm 0.68$ km s$^{-1}$ found by us previously in \mbox{\citet{golubov13}}. 
Rather than taking an old value of $V_{\odot}$ together with the radial scalelengths for three metallicity populations,
we re-determine them here 
in order to quantify the impact of our improved analysis in combination with the new distances and improved proper motions. 
To find the tangential component $V_{\odot}$ we apply to the local RAVE data the new Str\"omberg relation as in \mbox{\citet{golubov13}}, but now 
including the vertical correction terms discussed in the previous section.
We assume a Galactocentric distance of the Sun of \mbox{$R_0=8$ kpc}, which is consistent with \mbox{\citet{reid93}} 
as well as with a more recent study by \mbox{\citet{gillessen09}}. 
Then adopting the proper motion of Sgr A* from \mbox{\citet{reid05}} we get the solar Galactocentric velocity \mbox{$v_\odot=241.6 \, \pm 15$ km s$^{-1}$}.

Applying  Eq. \ref{JE2} at $R=R_0$ and using \mbox{$v_0 := v_c(R_0)  = v_{\odot} - V_{\odot} $} and \mbox{$\Delta V = v_\odot - \overline{v_{\phi}}$}, 
we write for the left-hand side:
\begin{eqnarray}
v_0^2 - \overline{v_{\phi}}^2 = (v_{\odot} - V_{\odot})^2 - (v_{\odot}- \Delta V)^2 \nonumber \\
= - \Delta V^2 + 2 v_{\odot}\Delta V - 2 v_{\odot}V_{\odot} + V_{\odot}^2.
\label{lhs2}
\end{eqnarray}
This leads to the new version of the Str\"{o}mberg relation:
\begin{equation}
V^{\prime} = V_{\odot} - \frac{V_{\odot}^2}{2 v_{\odot}} + \frac{\sigma_R^2}{k^{\prime}}
\quad\mbox{with}\quad k^{\prime} =  \frac{2 R_E}{R_0}v_{\odot}
\label{stromberg}
\end{equation}
and the generalized version of $V^{\prime}$:
\begin{eqnarray}
&& V^{\prime}:=\Delta V + \label{Vprime}\\ 
&& \frac{\sigma_R^2 -\sigma_{\phi}^2 +R_0 F(R_0,z) + \eta (\sigma_R^2 - \sigma_z^2)\left[ 1 - \frac{z}{h_{\nu\sigma}}\right]  -\Delta V^2}{2 v_{\odot}}.\nonumber
\end{eqnarray}

\begin{figure}
\centerline{\resizebox{1.\hsize}{!}{\includegraphics{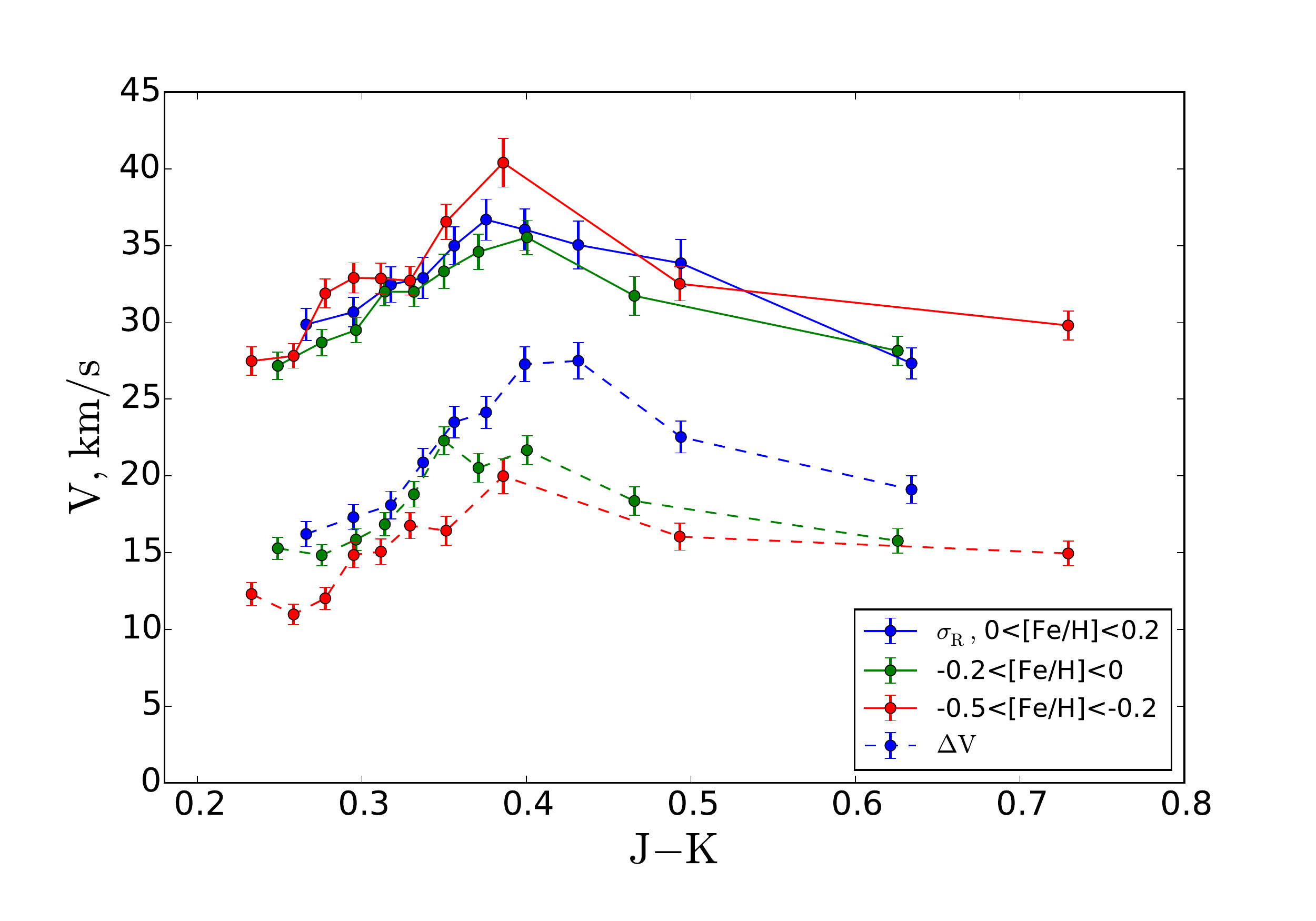}}}
\centerline{\resizebox{1.\hsize}{!}{\includegraphics{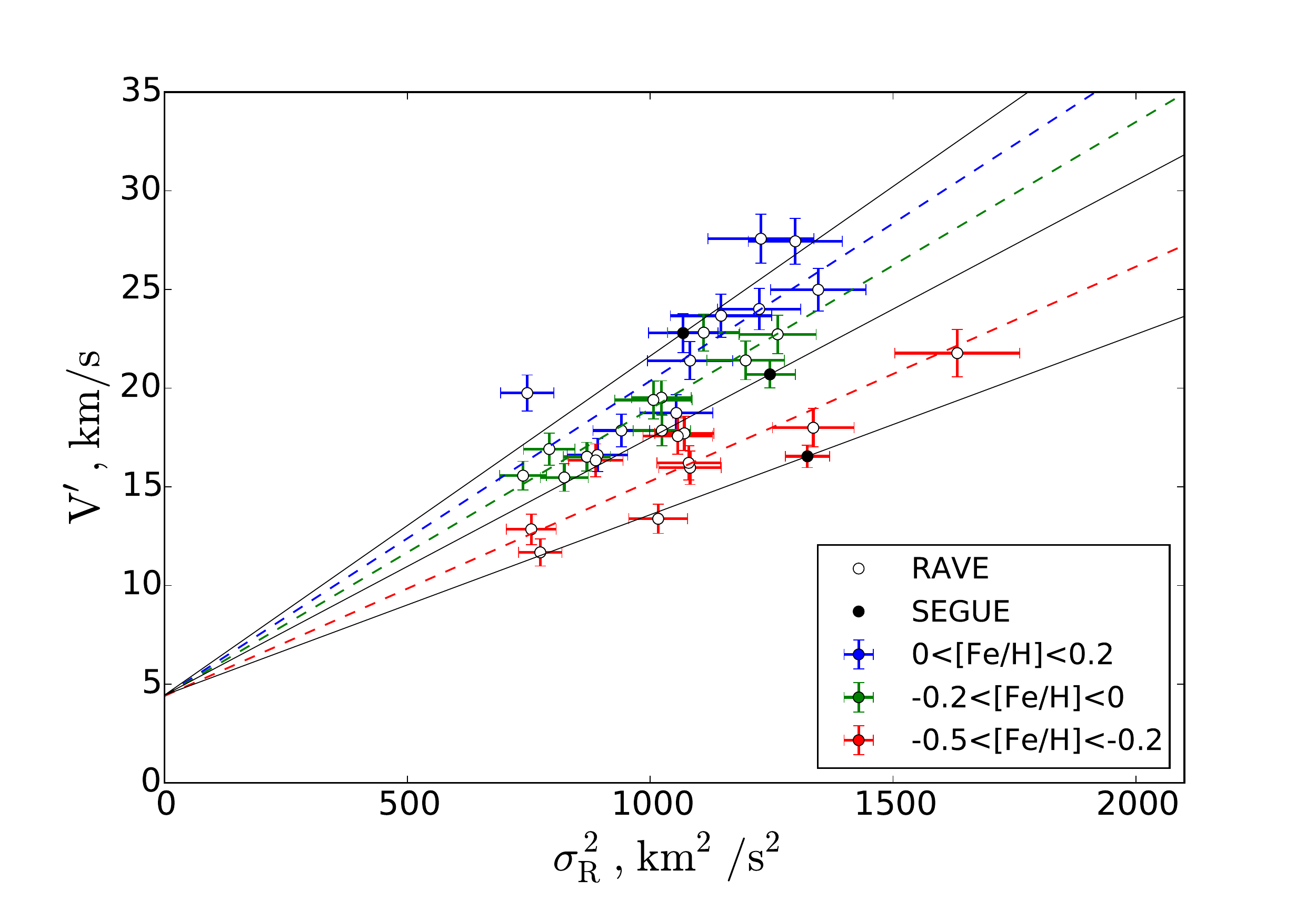}}}
\caption{The recalculation and consistency check of the asymmetric drift correction for the RAVE and local SEGUE data samples. 
The RAVE data are recalculated using the improved distances from \mbox{\citet{McMillan17}} and UCAC5 proper motions. 
The top panel shows $\sigma_\mathrm{R}$ 
and $\Delta V$ as functions of (J-K) 2MASS colour for different metallicity bins (compare to Figure 2 in \mbox{\citealp{golubov13}}). 
The bottom panel shows $V'$ as function of $\sigma_\mathrm{R}^2$ in view of the new Str\"omberg relation 
(Eqs. \ref{stromberg} and \ref{Vprime}) for each metallicity bin. The data points for the local G-dwarfs of SEGUE are added as full circles.
The metallicity binning for RAVE and SEGUE data is identical and has the same colour coding.
Only stars with $7.5\,\mathrm{kpc}<R<8.5\,\mathrm{kpc}$ and in the case of SEGUE with $|z|<1.5$ kpc are selected for the plot.
Dashed colour-coded lines are the linear least squares fit to the RAVE data. Black solid lines are added to readout the 
radial scalelengths corresponding to the positions of the SEGUE points and the value of $V_\odot$ determined from RAVE. 
Here and later on the error bars are calculated using the observational errors.
}
\label{stromberg-plot}
\end{figure}

To make practical use of Eqs. \ref{stromberg} and \ref{Vprime}, i.e., to determine values of $V_\odot$ and $R_E$, we need to bin our data sample 
into sub-bins with different kinematics. For this purpose we separate the RAVE sample in three subpopulations with different metallicities and then bin each subpopulation in 
(J-K) colour. The top panel of Figure \ref{stromberg-plot} shows such a binning of the squared radial velocity dispersion $\sigma_R^2$ and 
measured velocity $\Delta V$, which can be calculated straightforwardly, without knowledge of $V_\odot$. 
One can clearly see that the kinematic properties change systematically with both metallicity and colour.
Using the same binning we plot $V^{\prime}$ versus $\sigma_R^2$ (Figure \ref{stromberg-plot}, bottom panel). 
For simplicity, we assume here that $\eta=const.=0.8$ as derived in \mbox{\citet{binney14}} from the RAVE data. 
We also treat $h_{\nu\sigma}$ as independent of colour, though differing with metallicity and
select appropriate values from the disc model of \mbox{\citet{just10}} (see Appendix \ref{tracer}). 
The height above the midplane $z$ in Eq. \ref{Vprime} is calculated for each metallicity-colour bin as a mean value of the absolute $z$ for individual stars.

As in \mbox{\citet{golubov13}} we still see the systematic difference of $V'(\sigma_{R}^2)$ between metallicity bins, 
but the linear dependences show a larger scatter. The assumption that $R_E$, similar to $h_{\nu\sigma}$, is approximately the same for all colours inside
a given metallicity bin, but differs with metallicity, allows us to derive $R_E$ values for the three metallicity subpopulations as well as $V_{\odot}$. 
To do so we perform a simultaneous linear fit of the metallicity-colour sequences (shown with colour-coded dashed lines in Figure \ref{stromberg-plot}, bottom panel). 
The inverse slopes of the fitting lines $k^\prime_i$ can be directly converted to $R_{E,i}$, which under the assumption of constancy of the disc thickness and the shape of velocity ellipsoid
(see Section \ref{Jeans}) gives us the radial scalelengths for the selected populations. The solar peculiar motion can be readout from the $V^{\prime}$ extrapolated to zero 
radial velocity dispersion:
\begin{eqnarray}
R_{d,i} &=& \frac{R_0 k^\prime_i}{v_{\odot}}\\ \nonumber
V_{\odot} &=&  v_{\odot} - \sqrt{v_{\odot}^2 - 2 v_{\odot} \left. V^{\prime}\right|_{\sigma_R^2=0}} \, .
\end{eqnarray}

The updated value of the tangential component of the solar peculiar motion is found to be $V_\odot=4.47\pm 0.8$ km s$^{-1}$,
which translates to the local circular velocity $v_0 \approx 237 \pm 16$ \mbox{km s$^{-1}$}.
The radial scalelengths for the metallicity bins are
$R_d$(0$<$[Fe/H]$<$0.2$)=2.07\pm0.2$ kpc, $R_d$(-0.2$<$[Fe/H]$<$0$)=2.28\pm0.26$ kpc and $R_d$(-0.5$<$[Fe/H]$<$-0.2$)=3.05\pm0.43$ kpc,
which together with $V_\odot$ is in agreement with our old values from \mbox{\citet{golubov13}}.

\ifx
As an attentive reader may notice, one of the terms in $V'$ depends on the solar peculiar motion itself: 
$R_0 F(R_0,z)$ includes $v_0$, which depends on $V_{\odot}$ (consider the special case of Eq. \ref{JE2} for $R=R_0$ and the definition of $v_0$). 
However, changing the assumed value of $V_\odot$ by 3-4 km s$^{-1}$ influences the results only in the second 
decimal place, so we do not need to approach this in an iterative way. To be consistent with our findings, we use \mbox{$V_\odot=4.45\pm 0.8$ km s$^{-1}$} 
also for the vertical correction term. 
\fi

Now we must check whether SEGUE stars also follow the same trend.
For this purpose we split the thin disc SEGUE sample into the same metallicity bins and consider only stars with Galactocentric distances $7.5<R<8.5$ kpc.
The subdivision in colours is not possible for this data set because of its narrow colour range. 
Another characteristic of this sample is that G-dwarfs are distributed over a significantly larger range of $|z|$ (Figure \ref{samples-plot}, top). 
And though we do not reconstruct here the shape and orientation of the velocity ellipsoid as a function of $R$ and $z$, 
we may roughly account for the vertical gradients of $\sigma_R^2$ and $\sigma_z^2$. 
To do so we apply Eq. \ref{Vprime} for the calculation of $V'$ not to the whole metallicity bin, but separately in vertical sub-bins ($|z|=0 \, ... \, 1.5$ kpc with a step of 0.5 kpc) 
and calculate the resulting $V'$ by taking a weighted mean of the values obtained for different $|z|$.

The SEGUE points (filled circles in Figure \ref{stromberg-plot}, bottom panel) demonstrate good consistency with the RAVE data, which means that for
further analysis of SEGUE thin disc sample we can securely use the values of scalelengths derived from RAVE. 
This is an important result, accounting for all uncertainties of the metallicity calibration in RAVE and possible velocity biases for SEGUE.
To be even more precise we can inverse the problem and read out scalelengths for individual points adopting the solar peculiar velocity derived with the RAVE data. 
The values of $R_d$ calculated for the SEGUE data in such a way (see the three solid black lines in Figure \ref{stromberg-plot}, bottom panel) are: 
$R_d$(0$<$[Fe/H]$<$0.2$)=1.91\pm0.23$ kpc, $R_d$(-0.2$<$[Fe/H]$<$0$)=2.51\pm0.25$ kpc and $R_d$(-0.5$<$[Fe/H]$<$-0.2$)=3.55\pm0.42$ kpc.
We use these new values and $V_\odot=4.47\pm 0.8$ km s$^{-1}$ in our further analysis. 

The linearity of the asymmetric drift correction, i.e., constancy of $R_d$ versus $\sigma_R^2$ is still under debate.
Nevertheless, even if the asymmetric drift dependence on $\sigma_R^2$ in fact turns out to be nonlinear for small $\sigma_R^2$
(as assumed by \mbox{\citealp{schoenrich10}}),
it will correspond to some shift in the measured circular velocity,
but we do not expect this shift to change drastically at different Galactocentric radii.
To the first approximation this would produce only a parallel displacement of the measured rotation curve.
Being interested in the general shape of the rotation curve, not in the exact value of the rotation velocity,
we apply the solar velocity and the radial scalelengths derived by Eq. \ref{stromberg} at all Galactocentric radii.

\section{Asymmetric drift and rotation curve\label{RC}}

Now, with updated values for the solar peculiar velocity and the radial scalelengths for three populations of the selected metallicities,
we proceed to the determination of the rotation curve in the extended solar neighbourhood. 
We go back to the Jeans equation formulated for arbitrary $(R,z)$ (see Eq. \ref{JE2}) and apply it to the thin disc SEGUE stars. 
In principle, Eq. \ref{JE2} can be directly used for the determination of the rotation velocity as all terms on the right-hand side
are now known and the mean tangential speed can be expressed as $\overline{v_{\phi}} = v_\odot - \Delta V$, 
where $v_\odot$ is the tangential speed of the Sun and $(-\Delta V)$ is the mean observed tangential velocity with respect to the Sun. 
However, we would prefer to formulate the expression of the rotation velocity in terms of the asymmetric drift $V_a$, 
so we recall its definition:

\begin{equation}
V_a = v_c - \overline{v_{\phi}} = v_c(R) - v_\odot + \Delta V.
\label{AD_def}
\end{equation}

Inserting $\overline{v_{\phi}} = v_c - V_a$ into the left-hand side of Eq. \ref{JE2}, we arrive at the quadratic equation for the asymmetric drift:
\begin{eqnarray}
V_a^2 &-&2 v_c V_a  - R \Bigg( F(R,z) + \Bigg. \label{JE_general}\\
&& \Bigg. \eta \frac{\sigma_R^2 - \sigma_z^2}{R}\left[ 1 - \frac{z}{h_{\nu \sigma}}\right] 
- \frac{\sigma_R^2}{R_E} 
+ \frac{\sigma_R^2 -\sigma_{\phi}^2}{R}
\Bigg ) = 0. \nonumber
\end{eqnarray}
We solve Eq. \ref{JE_general} with respect to $V_a$ and get:
\begin{eqnarray}
V_a &=& v_c(R) - \Bigg\{ v_c^2(R) + R \bigg( F(R,z) + \phantom{\bigg(} \Bigg.\label{Va_root}\\
\Bigg.  \phantom{\bigg(} && \eta \frac{\sigma_R^2 - \sigma_z^2}{R}\left[ 1 - \frac{z}{h_{\nu \sigma}}\right] 
- \frac{\sigma_R^2}{R_E} 
+ \frac{\sigma_R^2 -\sigma_{\phi}^2}{R}
 \bigg) \Bigg\}^{1/2} \approx \nonumber
\end{eqnarray}
$$
\frac{-R F(R,z) - \eta (\sigma_R^2 - \sigma_z^2)\left[ 1 - \frac{z}{h_{\nu \sigma}}\right] 
+ \sigma_R^2\frac{R}{R_E} 
- (\sigma_R^2 -\sigma_{\phi}^2)
}{2v_c(R)}, 
$$
where the last line corresponds to the linear approximation ignoring the $V_a^2$ term. 
The difference of the non-linear and linear values for $V_a$ is of the order of 5\% or 1\,km s$^{-1}$.

The final formula for calculating the rotation velocity at radius R for each bin $(R,z)$ is from Eq. \ref{AD_def}
\begin{equation}
v_c(R) = \overline{v_{\phi}} + V_a = v_\odot - \Delta V + V_a
\label{Vc}
\end{equation}
with the asymmetric drift correction $V_a$ given by Eq. \ref{Va_root}. 
As $V_a$ is itself a function of $v_c(R)$, the determination of the
rotation velocity is an iterative procedure, during which we assume $v_c \propto R^{\alpha}$. 
We start with some small $\alpha$ as initial value, at each step of the iteration the reconstructed rotation curve is fitted and the new value of $\alpha$
is derived to be plugged back in Eq. \ref{Va_root} via $v_c(R)$. The iteration procedure converges very quickly, after two or three cycles.

\begin{figure}
\centerline{\resizebox{1\hsize}{!}{\includegraphics{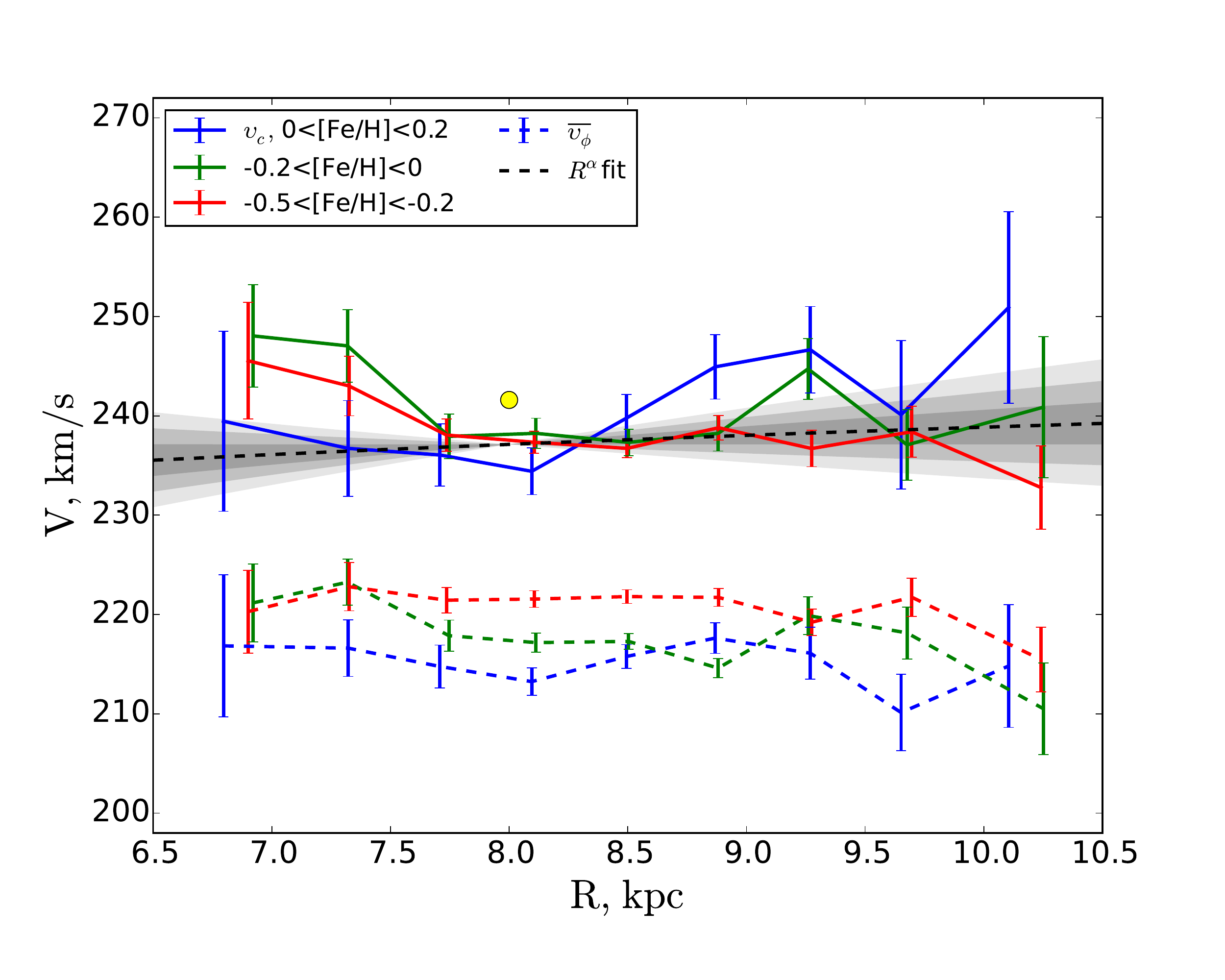}}}
\caption{The rotation curve in the extended solar neighbourhood traced via SEGUE stars.
The thin disc stars are split into the same three metallicity bins as before.
For each distance bin the mean rotation velocity $\overline{v_{\phi}}=v_{\odot}-\Delta V$ is measured (dashed curves, 
additional binning in $|z|$ applied in the range of $7\,\mathrm{kpc}<R<9\,\mathrm{kpc}$ is not shown here).
The circular velocity (solid curves) is calculated for the three metallicity bins.
The best power-law fit of the form $v_c=v_0(R/R_0)^{\alpha}$ is shown with a black dashed line.
The areas of the \mbox{1, 2 and 3$\sigma$-deviation} 
are shown with increasingly lighter shades of grey. 
The solar tangential velocity $v_\odot$ is marked with a yellow circle.}
\label{RC-drift-plot}
\end{figure}

\begin{figure}
\centerline{\resizebox{1\hsize}{!}{\includegraphics{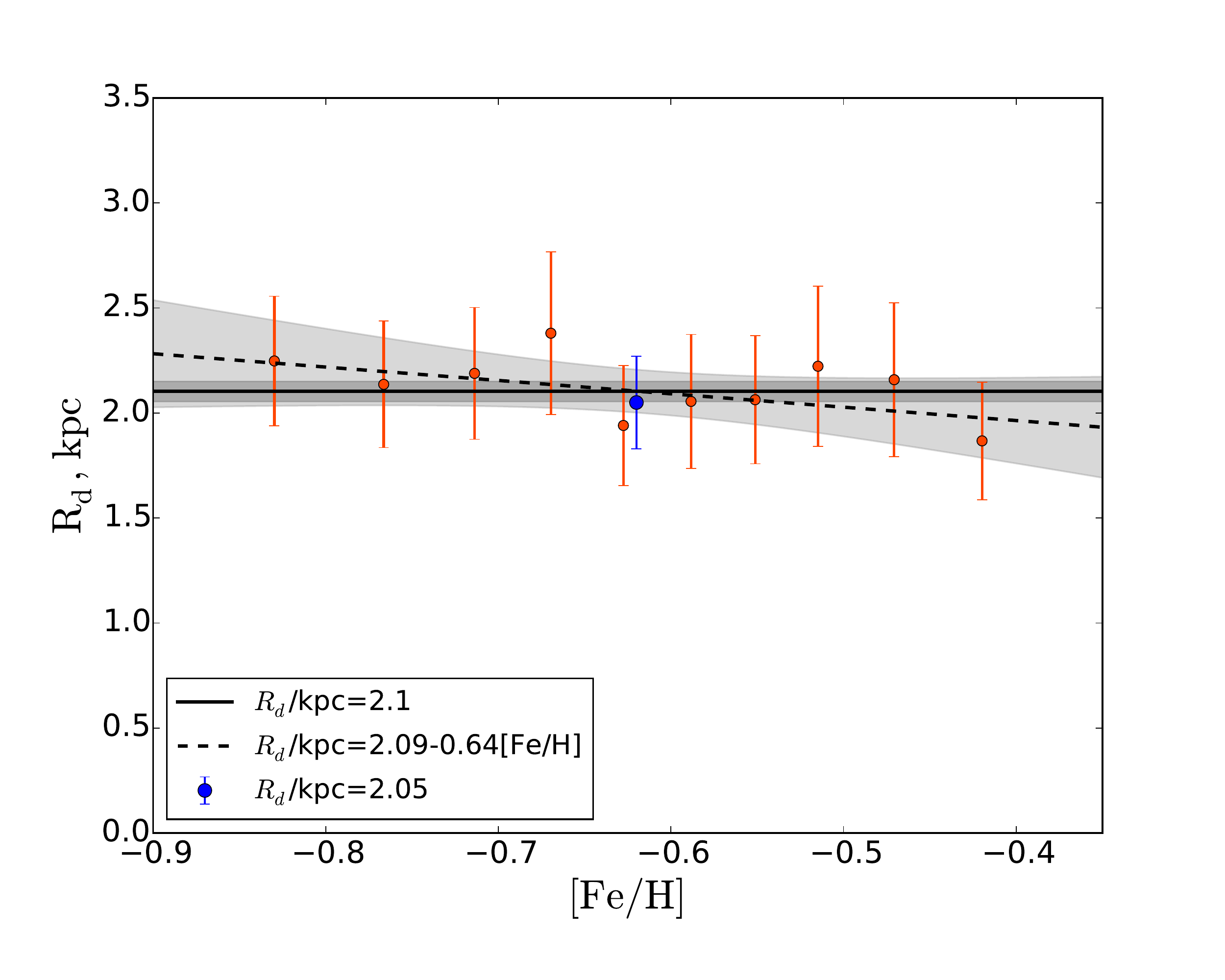}}}
\caption{The radial scalelength $R_\mathrm{d}$ of the thick disc versus metallicity.
Considered are SEGUE thick disc stars with $7.5\,\mathrm{kpc}<R<8.5\,\mathrm{kpc}$, 4195 in total. 
Red points show the radial scalelengths calculated for each metallicity bin with the 'effective' quantities in Eq. \ref{Vprime}.
The points are fitted with a constant and also with $R_d$ linearly dependent on [Fe/H] (solid and dashed lines).
One-sigma areas are shown in grey. The blue point corresponds to the value of scalelength derived in case of binning the same sample 
in 4 $|z|$-bins. 
}
\label{Rd-plot}
\end{figure}

We bin thin disc SEGUE stars, again separated in three metallicities as before, in Galactocentric distances with equal step of 0.4 kpc. 
The further data analysis is not the same for all distance bins. As we mentioned before,
the SEGUE stars are in general distributed over a large range in $|z|$, so we would like to take into account 
the vertical gradients in $\sigma_R^2$ and $\sigma_z^2$ by applying our equations separately at different heights for each distance. 
As one can see from the top panel of Figure \mbox{\ref{samples-plot}}, such $|z|$-binning for the thin disc sample is justified only close to $R_0$, 
approximately for 7\,kpc $<R<$ 9\,kpc, because at larger 
Galactocentric distances the majority of stars is located at approximately the same height, such that low-$|z|$ bins would be essentially empty and suffering
from high Poisson noise.
For this reason we do not bin in $|z|$ outside this distance range. Taking this into account, at $R<7$ kpc and $R>9$ kpc 
for each R-bin we find the mean tangential velocity $\overline{v_\phi}$ (dashed colour-coded lines in Figure~\ref{RC-drift-plot}), 
as well as three velocity dispersions and mean values of $R$ and $|z|$.
Then we apply the correction for the asymmetric drift from Eq. \ref{Va_root}, which is typically $\sim$20 \mbox{km s$^{-1}$}.
For $7\,\mathrm{kpc}<R<9\,\mathrm{kpc}$ we do the same, but separately for three vertical bins (as before, $|z|=0 \, ... \, 1.5$ kpc with a step of 0.5 kpc), 
and then calculate the weighted mean $v_c$ at each R. 
The obtained rotation curves representing different metallicity bins are shown as solid lines in Figure~\ref{RC-drift-plot} and are broadly consistent with each other 
within the error bars.

Between 8 and 10 kpc the rotation curve is flat to an accuracy of a few kilometres per second, 
and definitely does not show the 10\,km s$^{-1}$ dip described by \mbox{\citet{sofue,sofue10}}.
On the other hand, the inner part of our rotation curve demonstrates a metallicity-dependent rise with the amplitude up to \mbox{10 km s$^{-1}$},
similar to the one attested by \mbox{\citet{sofue,sofue10}}.

We perform a power-law fit $V_c\propto R^\alpha$ simultaneously to all three rotation curves and find a small power law index $\alpha=0.033\pm 0.034$ 
(fit shown in Figure~\ref{RC-drift-plot} with a dashed line).
This transforms into the local slope of the rotation curve $dV_c/dR=0.98\pm 1$ \mbox{km s$^{-1}$ kpc$^{-1}$}.
However, the existing measurements of the Oort constants point at a moderately negative slope: 
the classical value from \mbox{\citet{binney}} is $dV_c/dR=-2.4 \pm 1$ \mbox{km s$^{-1}$ kpc$^{-1}$}, 
while the more recent study by \mbox{\citet{bovy17}} from TGAS data suggests an even steeper slope of $-3.4 \pm 0.6$ \mbox{km s$^{-1}$ kpc$^{-1}$}.
Still, we must keep in mind that the Oort constants measure only a very local slope of the rotation curve, which may be stronger perturbed by the local spiral arm structure 
and the bar, while our analysis goes all the way to 1-2 kpc away from the Sun. 

The thick disc SEGUE stars are not very instrumental in reconstructing the rotation curve,
as the radial scalelength of the thick disc is poorly constrained.
We can solve the inverse problem: use the data to reconstruct the radial scalelength of the thick disc assuming the rotation curve already known. 
We express $R_\mathrm{d}$ from Eq. \ref{stromberg} and calculate it for ten metallicity bins of the thick disc. 
For the parameter $h_{\nu\sigma}$ of the thick disc we use a value of 800 pc, similar to its scaleheight \mbox{\citep{just11}}. 
Furthermore, we take only the local thick disc sample with $7.5\,\mathrm{kpc}<R<8.5\,\mathrm{kpc}$ in order to 
avoid regions, where our results can be biased by the uncertainty in the vertical correction term 
(this effect is not so important for the thin disc sample, as its stars are in general located at smaller $|z|$ than those of the thick disc).
The resulting $R_\mathrm{d}$ is shown in Figure~\ref{Rd-plot} with red points. 
Though the data points show some small variation of $R_d$ with [Fe/H], the constant value of the scalelength \mbox{$2.1 \pm 0.05$ kpc} is more robust 
(see the darker one-sigma area in Figure \ref{Rd-plot}). In other words, the chemically defined thick disc behaves as a single kinematically homogeneous population.
The value of the thick disc scalelength found here is consistent with 
simulations by \mbox{\citet{minchev15}} and with the data analysis by \mbox{\citet{bovy12}} and \mbox{\citet{kordopatis17}}.
When binning the local SEGUE thick disc sample in metallicity, we calculate $R_d$ for the 'effective' quantities in the bin: velocity dispersions determined for all stars
and also mean $|z|$ and R. Such 'effective' values are quite representative in case of R and velocity dispersions, because our data are very local in terms of Galactocentric distances
and expected to be kinematically homogeneous, so the velocity dispersions should not have a strong gradient in vertical direction. On the other hand, the mean value 
of $|z|$ might be misleading as the vertical distribution of stars is quite inhomogeneous. The vertical correction and the tilt term in $V'$ (Eq. \ref{Vprime}) 
are quite sensitive to $z$, so to cross-check our results we apply Eq. \ref{stromberg} for 4 $|z|$-bins (0...2 kpc with a 0.5 kpc step) with no binning in metallicity as 
the data are not abundant enough to allow a simultaneous separation in both $|z|$ and [Fe/H]. The resulting scalelength, which is calculated again as a weighted mean 
of the values found for different $|z|$, deviates only slightly from the value
found at the previous step: \mbox{$R_d = 2.05 \pm 0.22$ kpc}. We overplot it in Figure \ref{Rd-plot} with a blue dot for the mean metallicity of the local thick disc sample. 
A good agreement between the values of scalelength calculated via the different binning of our sample in the parameter space ensures us of the robustness of the derived result.

\section{Discussion and conclusions\label{conclusions}}

In this paper we performed the revision and improvement of the methods developed previously in \mbox{\citet{golubov13}}. 
Starting from the classical Jeans analysis we arrived at the new Str\"omberg relation for the asymmetric drift 
and applied it locally to the most recent RAVE data. This enabled us to update the value of the solar peculiar motion to 
$V_\odot=4.47\pm 0.8$ \mbox{km s$^{-1}$}. This is lower than the typical values reported by other authors, which are around 10-12 \mbox{km s$^{-1}$}. 
However, in the study of \mbox{\cite{sharma14}} also based on the RAVE data, 
though in the framework of a different Galaxy model, the solar peculiar velocity is smaller as well, $V_\odot=7.62^{+0.13}_{-0.16}$ km s$^{-1}$.
We also found radial scalelengths for the three metallicity populations, which are 
$R_d$(0$<$[Fe/H]$<$0.2$)=2.07\pm0.2$ kpc, $R_d$(-0.2$<$[Fe/H]$<$0$)=2.28\pm0.26$ kpc and $R_d$(-0.5$<$[Fe/H]$<$-0.2$)=3.05\pm0.43$ kpc. 
Our analysis demonstrates good consistency of the SEGUE and the RAVE data in terms of kinematics. With the peculiar velocity of the Sun derived from the RAVE sample, 
SEGUE data give similar values for the scalelengths, 
$R_d$(0$<$[Fe/H]$<$0.2$)=1.91\pm0.23$ kpc, $R_d$(-0.2$<$[Fe/H]$<$0$)=2.51\pm0.25$ kpc and $R_d$(-0.5$<$[Fe/H]$<$-0.2$)=3.55\pm0.42$ kpc.

Then we used the SEGUE sample of the thin disc G-dwarfs to reconstruct the rotation curve of the Milky Way, 
ranging from 7 to 10 kpc in Galactocentric radius. 
We took into account the asymmetric drift correction (Eq. \ref{Va_root}) and showed that 
the resulting rotation curve is essentially flat (Figure~\ref{RC-drift-plot}). 
Thus, the existence of any features in the rotation curve just outside the solar radius is discarded in the framework of our analysis.
The formal power-law fit to the rotation curve implies a positive slope $\alpha=0.033\pm 0.034$ consistent with a flat rotation curve, 
although we see that its local value is probably smaller. The corresponding 
radial gradient of the circular speed is $dV_c/dR=0.98\pm 1$ \mbox{km s$^{-1}$ kpc$^{-1}$}, which is 
in agreement with the findings of \mbox{\cite{sharma14}}, who derived a similar value from the RAVE data: 
$dV_c/dR=0.67^{+0.25}_{-0.26}$ \mbox{km s$^{-1}$ kpc$^{-1}$}.

Using SEGUE data and relying on the determined slope of the rotation curve, we also calculated the radial scalelength of the thick disc. 
It is $2.05\pm0.22$ kpc, and no strong dependence on metallicity was observed. 
Values of the quantities derived in this paper are summarized in Table \ref{tab-rez}.

Finally, we have to discuss the dependence of our results on the assumed constants and parameters. 

The pair ($R_0$,$v_\odot$), on the one hand, influences the derived stellar spatial distribution and velocities from the observables. 
On the other hand, it enters the equation for the asymmetric drift correction (see Eq. \ref{Va_root}).  
For the recommended values from \mbox{\citet{bhawthorn}} ($R_0$,$v_\odot$)$=$\mbox{(8.2 kpc, 248 km s$^{-1}$)}
the change in the re-calculated solar peculiar velocity and scaleheights for the three metallicity bins
lie well within one sigma and the changes in the rotation curve can be mostly described in terms of a
vertical shift to higher velocities and horizontal translation to larger Galactocentric distances. Its slope is in this case
$\alpha=0.024\pm 0.031$, which is again consistent with a flat rotation curve.

\begin{table}
\caption{The summary of the results.} 
\begin{tabular}{l|l|l} \hline
Quantity & from RAVE & from SEGUE \\ \hline
$V_\odot$ (km s$^{-1}$) & $4.47\pm 0.8$  &  \\
$R_d^\mathrm{thin}$(0$<$[Fe/H]$<$0.2$)$ (kpc) & $2.07\pm0.2$ & $1.91\pm0.23$ \\
$R_d^\mathrm{thin}$(-0.2$<$[Fe/H]$<$0$)$ (kpc) & $2.28\pm0.26$  & $2.51\pm0.25$ \\
$R_d^\mathrm{thin}$(-0.5$<$[Fe/H]$<$-0.2$)$ (kpc) & $3.05\pm0.43$ & $3.55\pm0.42$ \\
$\alpha$ &  & $0.033\pm 0.034$ \\
$R_d^\mathrm{thick}$ (kpc) &  & 2.05 $\pm$ 0.22\\ 
\hline
\end{tabular}
\label{tab-rez}
\end{table}

What is the impact of the solar peculiar motion?
Changes of $(dU,dV,dW)_\odot$ add quadratically to the corresponding velocity dispersions.
The vertical velocity has no other impact on the result, but we should check that the vertical component of the mean measured relative velocity 
of the stars in the sample is approximately $-W_\odot$, which is indeed the case. 
$U_\odot$ enters the velocity transformations to cylindrical coordinates, so every time 
we change $V_{\odot}$, $R_0$ or $v_\odot$, we have to adapt it to have 
$\overline{v_R} \approx 0$ as we assumed in Section \ref{Jeans}. However, this correction is small and we can neglect it
as it is surely beyond the accuracy we can hope to achieve in the framework of our approach. $V_{\odot}$ has a larger impact on the
results as the asymmetric drift correction depends on it, mainly via $v_c$ and the scalelengths (see Eq. \ref{Va_root}). 
We test two values of $V_{\odot}$, $\sim$ 3 km s$^{-1}$ \mbox{\citep{golubov13}} 
and $\sim$ 7.6 \mbox{km s$^{-1}$} \mbox{\citep{sharma14}}. The rotation curve slope is then 
$\alpha=0.039\pm 0.034$ and $\alpha=0.014\pm 0.028$, respectively. We also test the sensitivity of our results to the vertical 
scaleheights $h_{\nu\sigma}$ as they are not tightly constrained. Changing $h_{\nu\sigma}$ by $\pm20$\% leads to the slopes of
$\alpha=0.022\pm 0.03$ and $\alpha=0.049\pm 0.038$, which deviate from our standard value by less than 0.5$\sigma$.

To quantify the vertical gradient of the radial force term we assume a Galaxy model, i.e., use some form of the potential as an input. 
However, we believe that the rotation curve obtained in Section \ref{RC} is not strongly predefined by this choice. 
The $RF(R,z)$ term is not a dominating one in the asymmetric drift correction, so with respect to the rotation velocity it is a first-order correction. 
The modification of the potential will enter $v_c$ as a second-order correction only,
and this already meets the limit of our accuracy. 
As we inferred by running the tests with the $GalPot$ code, the main contribution to the vertical correction of the radial force 
comes from the thin disc, thus its scalelength and scaleheight are the main sources of uncertainty in this term. Taking a $R_{d}$ value of 2 or 3 kpc
we arrive at a rotation curve, which is correspondingly slightly steeper ($\alpha = 0.041 \pm 0.041$) or flatter ($\alpha = 0.027 \pm 0.029$) than the one 
we presented in Figure \ref{RC-drift-plot}. Testing $h_z$ of 200 and 400 pc results in similar changes: 
$\alpha = 0.041 \pm 0.036$ and $\alpha = 0.026 \pm 0.03$. 
The impact of varying $R_d$ and $h_z$ of the other discs is negligible.

None of these deviations from our main result produce essential changes in the derived rotation curve shape. So we can conclude
that in the framework of the developed analysis our outcome is robust with respect to the small changes of our constants 
and the pre-choice of the Galactic potential. Our analysis of the local rotation curve does not support the existence of any special features in its shape 
like a significant dip at R = 9 kpc.

\section*{Acknowledgements}

This work was supported by Sonderforschungsbereich SFB 881 'The Milky Way
System' (subprojects A6 and A5) of the German Research Foundation (DFG).

Funding for RAVE has been provided by: the Australian Astronomical Observatory; 
the Leibniz-Institut fuer Astrophysik Potsdam (AIP); the Australian National University; 
the Australian Research Council; the French National Research Agency; 
the German Research Foundation (SPP 1177 and SFB 881); the European Research Council (ERC-StG 240271 Galactica); 
the Istituto Nazionale di Astrofisica at Padova; The Johns Hopkins University; 
the National Science Foundation of the USA (AST-0908326); the W. M. Keck foundation; 
the Macquarie University; the Netherlands Research School for Astronomy; 
the Natural Sciences and Engineering Research Council of Canada; 
the Slovenian Research Agency (P1-0188); the Swiss National Science Foundation; 
the Science \& Technology Facilities Council of the UK; Opticon; Strasbourg Observatory; 
and the Universities of Groningen, Heidelberg and Sydney.
The RAVE web site is at \url{https://www.rave-survey.org}.

Funding for SDSS-I and SDSS-II has been provided by the Alfred P. Sloan Foundation, the Participating Institutions,
the National Science Foundation, the U.S. Department of Energy, the National Aeronautics and Space Administration,
the Japanese Monbukagakusho, the Max Planck Society, and the
Higher Education Funding Council for England. The SDSS Web Site is \url{http://www.sdss.org}.

UCAC5 proper motions and the distances from \mbox{\cite{McMillan17}} were obtained with the use of  
the European Space Agency (ESA) mission Gaia (\url{https://www.cosmos.esa.int/gaia}), 
processed by the Gaia Data Processing and Analysis Consortium (DPAC, \url{https://www.cosmos.esa.int/web/gaia/dpac/consortium}). 
Funding for the DPAC has been provided by national institutions, in particular the institutions participating in the Gaia Multilateral Agreement.

The authors are very grateful to Young-Sun Lee and Timothy C. Beers for providing their SEGUE data sample for our analysis and for fruitful discussions.

We also thank the anonymous referee for the detailed suggestions, which improved the paper significantly.

\bibliographystyle{aa}
\bibliography{rc15}

\begin{thebibliography}{}

\bibitem[\protect\citeauthoryear{{Binney} et~al.}{{Binney}
  et~al.}{2014}]{binney14}
{Binney} J., {Burnett} B., {Kordopatis} G., et~al., 2014, \mnras, 439, 1231

\bibitem[\protect\citeauthoryear{{Binney} \& {Tremaine}}{{Binney} \&
  {Tremaine}}{2008}]{binney}
{Binney} J.,  {Tremaine} S., 2008, {Galactic Dynamics}.
\newblock Princeton Univ. Press

\bibitem[\protect\citeauthoryear{{Bland-Hawthorn} \&
  {Gerhard}}{{Bland-Hawthorn} \& {Gerhard}}{2016}]{bhawthorn}
{Bland-Hawthorn} J.,  {Gerhard} O., 2016, \araa, 54, 529

\bibitem[\protect\citeauthoryear{{Bovy}}{{Bovy}}{2017}]{bovy17}
{Bovy} J., 2017, \mnras, 468, L63

\bibitem[\protect\citeauthoryear{{Bovy} et~al.}{{Bovy} et~al.}{2012a}]{bovy12a}
{Bovy} J., {Allende Prieto} C., {Beers} T.~C., et~al., 2012a, \apj, 759, 131

\bibitem[\protect\citeauthoryear{{Bovy} et~al.}{{Bovy} et~al.}{2012b}]{bovy12}
{Bovy} J., {Rix} H.-W., {Liu} C., et~al., 2012b, \apj, 753, 148

\bibitem[\protect\citeauthoryear{{Chemin} et~al.}{{Chemin}
  et~al.}{2016}]{chemin16}
{Chemin} L., {Hur{\'e}} J.-M., {Soubiran} C., et~al., 2016, \aap, 588, A48

\bibitem[\protect\citeauthoryear{{Chemin}, {Renaud}, \& {Soubiran}}{{Chemin}
  et~al.}{2015}]{chemin15}
{Chemin} L., {Renaud} F.,  {Soubiran} C., 2015, \aap, 578, A14

\bibitem[\protect\citeauthoryear{{Cheng} et~al.}{{Cheng}
  et~al.}{2012}]{cheng12}
{Cheng} J.~Y., {Rockosi} C.~M., {Morrison} H.~L., et~al., 2012, \apj, 752, 51

\bibitem[\protect\citeauthoryear{{Dehnen}}{{Dehnen}}{2000}]{dehnen00}
{Dehnen} W., 2000, \aj, 119, 800

\bibitem[\protect\citeauthoryear{{Dehnen} \& {Binney}}{{Dehnen} \&
  {Binney}}{1998a}]{dehnen98a}
{Dehnen} W.,  {Binney} J., 1998a, \mnras, 294, 429

\bibitem[\protect\citeauthoryear{{Dehnen} \& {Binney}}{{Dehnen} \&
  {Binney}}{1998b}]{dehnen98}
{Dehnen} W.,  {Binney} J.~J., 1998b, \mnras, 298, 387

\bibitem[\protect\citeauthoryear{{Eisenstein} et~al.}{{Eisenstein}
  et~al.}{2011}]{eisenstein11}
{Eisenstein} D.~J., {Weinberg} D.~H., {Agol} E., et~al., 2011, \aj, 142, 72

\bibitem[\protect\citeauthoryear{{Gaia Collaboration} et~al.}{{Gaia
  Collaboration} et~al.}{2016}]{gaia16}
{Gaia Collaboration} , {Brown} A.~G.~A., {Vallenari} A., et~al., 2016, \aap,
  595, A2

\bibitem[\protect\citeauthoryear{{Gillessen} et~al.}{{Gillessen}
  et~al.}{2009}]{gillessen09}
{Gillessen} S., {Eisenhauer} F., {Fritz} T.~K., et~al., 2009, \apjl, 707, L114

\bibitem[\protect\citeauthoryear{{Golubov} et~al.}{{Golubov}
  et~al.}{2013}]{golubov13}
{Golubov} O., {Just} A., {Bienaym{\'e}} O., et~al., 2013, \aap, 557, A92

\bibitem[\protect\citeauthoryear{{Huang} et~al.}{{Huang}
  et~al.}{2016}]{huang16}
{Huang} Y., {Liu} X.-W., {Yuan} H.-B., et~al., 2016, \mnras, 463, 2623

\bibitem[\protect\citeauthoryear{{Just}, {Gao}, \& {Vidrih}}{{Just}
  et~al.}{2011}]{just11}
{Just} A., {Gao} S.,  {Vidrih} S., 2011, \mnras, 411, 2586

\bibitem[\protect\citeauthoryear{{Just} \& {Jahrei{\ss}}}{{Just} \&
  {Jahrei{\ss}}}{2010}]{just10}
{Just} A.,  {Jahrei{\ss}} H., 2010, \mnras, 402, 461

\bibitem[\protect\citeauthoryear{{Kafle} et~al.}{{Kafle}
  et~al.}{2012}]{kafle12}
{Kafle} P.~R., {Sharma} S., {Lewis} G.~F.,  {Bland-Hawthorn} J., 2012, \apj,
  761, 98

\bibitem[\protect\citeauthoryear{{Kordopatis} et~al.}{{Kordopatis}
  et~al.}{2017}]{kordopatis17}
{Kordopatis} G., {Wyse} R.~F.~G., {Chiappini} C., et~al., 2017, \mnras, 467,
  469

\bibitem[\protect\citeauthoryear{{Kunder} et~al.}{{Kunder}
  et~al.}{2017}]{kunder17}
{Kunder} A., {Kordopatis} G., {Steinmetz} M., et~al., 2017, \aj, 153, 75

\bibitem[\protect\citeauthoryear{{Lee} et~al.}{{Lee} et~al.}{2011}]{lee11}
{Lee} Y.~S., {Beers} T.~C., {An} D., et~al., 2011, \apj, 738, 187

\bibitem[\protect\citeauthoryear{{Liu} et~al.}{{Liu} et~al.}{2014}]{liu14}
{Liu} X.-W., {Yuan} H.-B., {Huo} Z.-Y., et~al., 2014, {LSS-GAC - A LAMOST
  Spectroscopic Survey of the Galactic Anti-center}, in IAU Symposium, Vol.
  298, {Feltzing} S., {Zhao} G., {Walton} N.~A.,  {Whitelock} P. (eds.),
  Setting the scene for Gaia and LAMOST, p. 310

\bibitem[\protect\citeauthoryear{{L{\'o}pez-Corredoira}}{{L{\'o}pez-Corredoira}}{2014}]{lopez14}
{L{\'o}pez-Corredoira} M., 2014, \aap, 563, A128

\bibitem[\protect\citeauthoryear{{McMillan} et~al.}{{McMillan}
  et~al.}{2017}]{McMillan17}
{McMillan} P.~J., {Kordopatis} G., {Kunder} A., et~al., 2017, ArXiv
  e-prints:[1707.04554]

\bibitem[\protect\citeauthoryear{{Minchev} et~al.}{{Minchev}
  et~al.}{2015}]{minchev15}
{Minchev} I., {Martig} M., {Streich} D., et~al., 2015, \apjl, 804, L9

\bibitem[\protect\citeauthoryear{{Monari}, {Famaey}, \& {Siebert}}{{Monari}
  et~al.}{2016}]{monari16}
{Monari} G., {Famaey} B.,  {Siebert} A., 2016, \mnras, 457, 2569

\bibitem[\protect\citeauthoryear{{Monari} et~al.}{{Monari}
  et~al.}{2017}]{monari17}
{Monari} G., {Kawata} D., {Hunt} J.~A.~S.,  {Famaey} B., 2017, \mnras, 466,
  L113

\bibitem[\protect\citeauthoryear{{Navarro}, {Frenk}, \& {White}}{{Navarro}
  et~al.}{1995}]{nfw}
{Navarro} J.~F., {Frenk} C.~S.,  {White} S.~D.~M., 1995, \mnras, 275, 720

\bibitem[\protect\citeauthoryear{{P{\'e}rez-Villegas}
  et~al.}{{P{\'e}rez-Villegas} et~al.}{2017}]{perez-villegas17}
{P{\'e}rez-Villegas} A., {Portail} M., {Wegg} C.,  {Gerhard} O., 2017, \apjl,
  840, L2

\bibitem[\protect\citeauthoryear{{Reid}}{{Reid}}{1993}]{reid93}
{Reid} M.~J., 1993, \araa, 31, 345

\bibitem[\protect\citeauthoryear{{Reid} \& {Brunthaler}}{{Reid} \&
  {Brunthaler}}{2005}]{reid05}
{Reid} M.~J.,  {Brunthaler} A., 2005, {The Proper Motion of Sgr A*}, in
  Astronomical Society of the Pacific Conference Series, Vol. 340, {Romney} J.,
   {Reid} M. (eds.), Future Directions in High Resolution Astronomy, p. 253

\bibitem[\protect\citeauthoryear{{Reid} et~al.}{{Reid} et~al.}{2014}]{reid14}
{Reid} M.~J., {Menten} K.~M., {Brunthaler} A., et~al., 2014, \apj, 783, 130

\bibitem[\protect\citeauthoryear{{Robin} et~al.}{{Robin} et~al.}{2003}]{robin3}
{Robin} A.~C., {Reyl{\'e}} C., {Derri{\`e}re} S.,  {Picaud} S., 2003, \aap,
  409, 523

\bibitem[\protect\citeauthoryear{{Sch{\"o}nrich}, {Binney}, \&
  {Dehnen}}{{Sch{\"o}nrich} et~al.}{2010}]{schoenrich10}
{Sch{\"o}nrich} R., {Binney} J.,  {Dehnen} W., 2010, \mnras, 403, 1829

\bibitem[\protect\citeauthoryear{{Sharma} et~al.}{{Sharma}
  et~al.}{2014}]{sharma14}
{Sharma} S., {Bland-Hawthorn} J., {Binney} J., et~al., 2014, \apj, 793, 51

\bibitem[\protect\citeauthoryear{{Siebert} et~al.}{{Siebert}
  et~al.}{2012}]{siebert12}
{Siebert} A., {Famaey} B., {Binney} J., et~al., 2012, \mnras, 425, 2335

\bibitem[\protect\citeauthoryear{{Skrutskie} et~al.}{{Skrutskie}
  et~al.}{2006}]{skrutskie06}
{Skrutskie} M.~F., {Cutri} R.~M., {Stiening} R., et~al., 2006, \aj, 131, 1163

\bibitem[\protect\citeauthoryear{{Sofue}, {Honma}, \& {Omodaka}}{{Sofue}
  et~al.}{2009}]{sofue}
{Sofue} Y., {Honma} M.,  {Omodaka} T., 2009, \pasj, 61, 227

\bibitem[\protect\citeauthoryear{{Sofue}, {Honma}, \& {Omodaka}}{{Sofue}
  et~al.}{2010}]{sofue10}
{Sofue} Y., {Honma} M.,  {Omodaka} T., 2010, \pasj, 62, 1367

\bibitem[\protect\citeauthoryear{{Steinmetz} et~al.}{{Steinmetz}
  et~al.}{2006}]{steinmetz06}
{Steinmetz} M., {Zwitter} T., {Siebert} A., et~al., 2006, \aj, 132, 1645

\bibitem[\protect\citeauthoryear{{Wegg}, {Gerhard}, \& {Portail}}{{Wegg}
  et~al.}{2015}]{wegg15}
{Wegg} C., {Gerhard} O.,  {Portail} M., 2015, \mnras, 450, 4050

\bibitem[\protect\citeauthoryear{{Wojno} et~al.}{{Wojno}
  et~al.}{2017}]{wojno17}
{Wojno} J., {Kordopatis} G., {Piffl} T., et~al., 2017, \mnras, 468, 3368

\bibitem[\protect\citeauthoryear{{Wojno} et~al.}{{Wojno}
  et~al.}{2016}]{wojno16}
{Wojno} J., {Kordopatis} G., {Steinmetz} M., et~al., 2016, \mnras, 461, 4246

\bibitem[\protect\citeauthoryear{{Yanny} et~al.}{{Yanny}
  et~al.}{2009}]{yanny09}
{Yanny} B., {Rockosi} C., {Newberg} H.~J., et~al., 2009, \aj, 137, 4377

\bibitem[\protect\citeauthoryear{{Zacharias}, {Finch}, \&
  {Frouard}}{{Zacharias} et~al.}{2017}]{zacharias17}
{Zacharias} N., {Finch} C.,  {Frouard} J., 2017, \aj, 153, 166

\end{thebibliography}

\begin{appendix}

\section{Vertical gradient of tracer populations\label{tracer}}

The tilt term in Eq. \ref{JE1} describes the vertical density gradient and the change in the tilt of the velocity ellipsoid. 
The tilt angle of the velocity ellipsoid is defined as
\begin{equation}
\frac{1}{2}\tan(2\alpha_{tilt}) = \frac{\sigma_{Rz}^2}{\sigma_R^2 - \sigma_z^2}
\label{tilt}
\end{equation}
On the other side, the tilt angle can be parametrized with a parameter $\eta$, which describes the orientation
of the velocity ellipsoid relative to the Galactic centre
\begin{equation}
\tan{(\alpha_{tilt})} = \eta \frac{z}{R}
\label{tiltparam}
\end{equation}
Assuming that $\alpha_{tilt}$ is small one then arrives at the following relation for $\sigma_{Rz}^2$:
\begin{equation}
\sigma_{Rz}^2 = \eta (\sigma_R^2 - \sigma_z^2) z/R.
\label{sig_rz1}
\end{equation}
The derivative of $\sigma_{Rz}^2$ is
\begin{eqnarray}
&&\frac{\partial \sigma_{Rz}^2}{\partial z} = \frac{\eta}{R}(\sigma_R^2-\sigma_z^2) - \frac{z}{R}(\sigma_R^2-\sigma_z^2)\frac{\partial \eta}{\partial z} + \label{srz_deriv}\\
&&\eta \frac{z}{R} \left(\frac{\partial \sigma_R^2}{\partial z} - \frac{\partial \sigma_z^2}{\partial z}\right)
= \sigma_{Rz}^2 \left[ \frac{1}{z} + \frac{\partial \ln \eta}{\partial z} + \frac{\partial \ln(\sigma_R^2-\sigma_z^2)}{\partial z} \right] \nonumber
\end{eqnarray}
Thus, the tilt term in Eq. \ref{JE1} will give together with Eq. \ref{srz_deriv}:
\begin{equation}
\sigma_{Rz}^2 \left[ \frac{1}{z} +  \frac{\partial \ln(\eta\nu (\sigma_R^2-\sigma_z^2))}{\partial z} \right].
\label{tilt1}
\end{equation}
The second  term in the brackets 
can be parametrized with some characteristic scaleheight $h_{\nu \sigma}$, so we arrive at
\begin{eqnarray}
\sigma_{Rz}^2 \frac{\partial \ln (\nu \sigma_{Rz}^2)}{\partial z} &=& \frac{\sigma_{Rz}^2}{z} \left[ 1 - \frac{z}{h_{\nu \sigma}}\right] = \nonumber\\
&& \eta \frac{\sigma_R^2 - \sigma_z^2}{R}\left[ 1 - \frac{z}{h_{\nu \sigma}}\right]
\label{terms3-4}
\end{eqnarray}

We expect that the vertical variation of $\eta$ and of $(\sigma_R^2-\sigma_z^2) $ is small compared to the gradient of the tracer density $\nu$. 
So we can relate the scaleheight $h_{\nu\sigma}$ to the characteristic scaleheight of the tracer density. 
For this purpose we use
the half-thickness values of the mono-age subpopulations with ages of 4, 6, and 10 Gyr, taken from the local thin disc model \mbox{\citep{just10}}.
The youngest subpopulation represents the bin with the highest metallicity. The adopted values of $h_{\nu\sigma}$ are 360, 430, and 530 pc, respectively.  

\end{appendix}

\end{document}